\documentclass[preprint,12pt]{elsarticle}
\usepackage[centertags]{amsmath}
\usepackage{amsfonts}
\usepackage{amssymb}
\usepackage{amsthm}
\usepackage{pgf}
\usepackage[T1]{fontenc}
\usepackage{indentfirst}
\usepackage{mathrsfs}
\usepackage[ansinew]{inputenc}
\usepackage{multirow}
\usepackage{hyperref}
\hypersetup{colorlinks=true, linkcolor=blue, citecolor=blue, filecolor=blue, urlcolor=blue,
            pdfproducer={Latex},
            pdfcreator={pdflatex}}

\journal{Physica A: Statistical Mechanichs and its Applications}

\begin{document}




\title{A deformed derivative model for turbulent diffusion of contaminants in the atmosphere}


\author[]{A. G. Goulart}

\author[]{M.~J. Lazo \corref{mycorrespondingauthor}}
\ead{matheuslazo@furg.br}
\cortext[mycorrespondingauthor]{Corresponding author}

\author[]{J. M. S. Suarez }

\address[]{Instituto de Matem\'atica, Estat\'{\i}stica e F\'{\i}sica, Universidade Federal do Rio Grande, 96.201-900 Rio Grande, Rio Grande do Sul, Brazil.}

\begin{abstract}

In the present work, we propose an advection-diffusion equation with Hausdorff deformed derivatives to stud the turbulent diffusion of contaminants in the atmosphere. We compare the performance of our model to fit experimental data against models with classical and Caputo fractional derivatives. We found that the Hausdorff equation gives better results than the tradition advection-diffusion equation when fitting experimental data. Most importantly, we show that our model and the Caputo fractional derivative model display a very similar performance for all experiments. This last result indicates that regardless of the kind of non-classical derivative we use, an advection-diffusion equation with non-classical derivative displaying power-law mean square displacement is more adequate to describe the diffusion of contaminants in the atmosphere than a model with classical derivatives. Furthermore, since Hausdorff derivatives can be related to several deformed operators, and since differential equations with the Hausdorff derivatives are easier to solve than equations with Caputo and other non-local fractional derivatives, our result highlights the potential of deformed derivative models to describe the diffusion of contaminants in the atmosphere.

\end{abstract}



\maketitle

\section{Introduction}

The dispersion of pollutants in the atmosphere is a permanent source of physical and mathematical challenging problems due to the turbulence of the airflow.
The multi-scaling behavior of a turbulent medium, related to its fractal structure \cite{Mandelbrot}, has as an important physical consequence the emergence of the anomalous diffusion phenomenon that plays a central role in the dispersion of contaminants in the Planetary Boundary Layer (PBL). A diffusion process is considered anomalous if it displays a non-linear mean square displacement, instead of a linear one that characterizes a common diffusive process \cite{Metzler2000}. Remarkably, although anomalous diffusion is found in a wide variety of real-world systems \cite{Metzler2000, Metzler2004}, it was first observed in nature in 1926 by Richardson when measuring the increase in the width of plumes of smoke from chimneys located in a turbulent velocity field \cite{Metzler2000, Metzler2004, Richardson, West,West2}.

The non-linear behavior of the increase in the width of plumes, found by Richardson, is a consequence of the fractal structure of the turbulent velocity field, where the fluctuation's size scales are much larger than the average scale \cite{Mandelbrot}.
Moreover, from this fractal behavior, it is expected that the classical advection-diffusion equation does not describe adequately the dispersion of pollutants in the atmosphere since the parameters of the system usually grow faster than the solution obtained from the classical advection-diffusion equation \cite{West,West2}. In this scenario, the customary strategy used to model the problem of dispersion of pollutants in the atmosphere is by using complex Eulerian and Lagrangian models \cite{Gryning1987, Hanna and Paine, Moreira2005a, Wilson}. While the Eulerian models are based on the advection-diffusion equation, the Lagrangian models are grounded on the  Langevin equation. In order to handle the anomalous diffusion, a usual approach is to modify the Eulerian models by considering that the turbulent flow and velocity fields can be modeled by a complex eddy diffusivity and mean velocity profile that is both considered as functions of spatial coordinates. These functions are set in order to fit experimental data or they are obtained from the Taylor statistical diffusion theory \cite{Taylor, Batchelor, Degrazia, Goulart2004}.

In recent works, we take a different route in order to deal with anomalous diffusion induced by turbulence. We show that an advection-diffusion equation with fractional derivatives arouses naturally in an anomalous diffusion process, displaying a power-law mean square displacement, describing a steady-state spatial concentration distribution of contaminants \cite{Goulart2017,Goulart2019}. The fractional calculus is a branch of mathematics that deals with non-integer order derivatives and integrals. Despite the fractional calculus emerged as a valuable mathematical tool to describe the time evolution of anomalous diffusion process in a wide variety of real-world systems \cite{Metzler2000, Metzler2004}, the use of fractional differential equations to describes the steady-state concentration of contaminants in the atmosphere is largely underexplored \cite{Goulart2017,Goulart2019}. In \cite{Goulart2017,Goulart2019} we showed that the advection-diffusion equation with Caputo fractional derivative, even with constant velocity field and a constant eddy diffusivity, gives very good results when fitting experimental data. Furthermore, the results presented in \cite{Goulart2019} indicate that should be a relationship between the order of the fractional derivative that better fits the experimental data with the physical structure of the turbulent flow.
However, since the Caputo fractional derivatives are non-local operators, it is not a simple task to obtain a direct relation of the order of the fractional derivative with the fractal structure of a turbulent flow.

The promising result found in our previous works \cite{Goulart2017,Goulart2019} motivates a deeper analysis of the relationship between turbulent flow and anomalous diffusion expressed mathematically by a diffusion equation with non-classical derivatives. In the present work, we propose a diffusion equation with a Hausdorff deformed derivatives to model the steady-state spatial concentration of contaminants in the atmosphere. The Hausdorff derivative is a kind of operator that can be called deformed derivatives \cite{WL,W}. Deformed derivatives includes the famous $q$-derivative \cite{qderivative} and several metric and local fractional derivatives \cite{WL,W}.
The Hausdorff derivative was introduced in the context of a time-dependent anomalous diffusion process \cite{Hausdorff} that displays a power-law mean square displacement. Since a fractional diffusion equation with Caputo derivatives also describes an anomalous process with power-law mean square displacement, we can ask which of the two derivatives would be more appropriate to model the diffusion of contaminants in the atmosphere? Furthermore, the Hausdorff derivatives have two advantages against Caputo in the analyses of a turbulent anomalous diffusion process. The first is that they are local operators, resulting in a diffusion equation easier to solve. The second advantage is that they can be easily related to the fractal dimension of the medium \cite{Balankin,Balankin2,Balankin3,Balankin4}.

Within this purpose, the main objective of the present work is to investigate the potential of a diffusion equation with Hausdorff derivative to fit real data for the steady-state spatial concentration of contaminants in the atmosphere. We compare the solution of the proposed model containing Hausdorff derivatives with the paradigmatic experiments of Copenhagen \cite{Gryning1984}, Prairie Grass \cite{Barad} and Hanford \cite{Nickola}. Furthermore, we also compare the present model with the solutions from the Caputo fractional derivative model \cite{Goulart2019}, and also to integer-order models. For constants velocity fields and eddy diffusivity, we found that the Hausdorff advection-diffusion equation we formulate gives better results than the tradition advection-diffusion equation when fitting experimental data. Most importantly, we show that our model and the Caputo fractional derivative model \cite{Goulart2019} display a very similar performance for all experiments. This last result indicates that regardless of the kind of non-classical derivative we use, an advection-diffusion equation with non-classical derivative displaying power-law mean square displacement is more adequate to describe the diffusion of contaminants in the atmosphere than a model with classical derivative. Furthermore, since Hausdorff derivatives can be related to several deformed derivatives \cite{WL,W}, and since differential equations with the Hausdorff derivatives are easier to solve than equations with Caputo and other non-local fractional derivatives, our result highlights the potential of deformed derivatives models to describe the diffusion of contaminants in the atmosphere.

The work is organized in the following way. In section \ref{SecH} we present the definition of Hausdorff derivative and its basic properties. The diffusion model we proposed is introduced in section \ref{SecM}. The result and analyses are displayed in section \ref{SecR}. Finally, the conclusions are presented in section \ref{SecC}.

\section{The Hausdorff and deformed derivatives}
\label{SecH}

In this section, we present the definition of some deformed derivatives and their relation with the Hausdorff derivative. A more detailed discussion, including the physical justifications for the relationship of the Hausdorff derivative with the fractal structure of the medium, is beyond the scope of this work but can be found in \cite{Hausdorff,Balankin,Balankin2,Balankin3,Balankin4}. Our main objective is to investigate the potential of a diffusion equation with deformed derivatives to fit real data for the steady-state spatial concentration of contaminants in the atmosphere. As the Hausdorff derivatives are related to several deformed derivatives, the results we obtained in the present work are general in the sense that our analyses can be extended to a huge variety of deformed derivatives.

\subsection{Hausdorff derivative}

The Hausdorff derivative was introduced by Chen \cite{Hausdorff} in the context of a  time-dependent anomalous diffusion process. Recently, it was successfully used to fit some real data for a fluid in a porous and fractured medium \cite{Balankin,Balankin2} and to study diffusion and random walk on fractals \cite{Balankin4}. The Hausdorff derivative of order $\alpha>0$ is defined as \cite{Hausdorff}:
\begin{equation}
\label{Hdef}
\frac{d}{dx^\alpha}f(x)= \lim_{x' \to x} \frac{f(x') - f(x)}{x'^\alpha - x^\alpha},
\end{equation}
where the parameter $\alpha$ can be related to the metric (mass) dimension $D<n$ of the fractal space contained in a Euclidean space $\mathbb{R}^n$, and to the fractal dimension $d_x$ of the intersection of the fractal with the Cartesian plane in $\mathbb{R}^n$ normal to $x$. This relation is given by relation $\alpha=D-d_x$ \cite{Balankin,Balankin2}.

If $f(x)$ is a differentiable function, the Hausdorff derivative \eqref{Hdef} can be related to an usual derivative by:
\begin{equation}
\label{Hdef2}
\frac{d}{dx^\alpha}f(x)= \lim_{x' \to x} \frac{f(x') - f(x)}{x'-x}\frac{x'-x}{x'^\alpha - x^\alpha}= \frac{x^{1-\alpha}}{\alpha}\frac{d}{dx}f(x).
\end{equation}
It is important to notice that, in this case, if $\alpha=1$ the Hausdorff derivative reduces to the classical first-order derivative.

Finally, the Hausdorff derivative \eqref{Hdef} can also be related with the Balankin's Hausdorff derivative \cite{Balankin,Balankin2}
\begin{equation}
\label{Hdef3}
\frac{D}{Dx^\alpha}f(x)= \left(\frac{x}{l_0}+1\right)^{1-\alpha}\frac{d}{dx}f(x)
\end{equation}
by the change of variable $x\rightarrow \frac{x}{l_0}+1$ in \eqref{Hdef2}, where $l_0$ is the fractal lower cutoff along the Cartesian axis $x$.

\subsection{conformable derivative}

The Hausdorff derivatives belong to a class of operators currently referred to as local fractional derivatives. For differentiable functions, apart from a multiplicative constant, it can be related to other local fractional derivatives as the conformable derivative \cite{comformable}
\begin{equation}
T_\alpha f(x)=\lim_{\epsilon \to 0} \frac{f(x+\epsilon x^{1-\alpha}) - f(x)}{\epsilon}, \label{talpha1}
\end{equation}
that gives $T_\alpha f(x)=x^{1-\alpha} f'(x)$ for differentiable functions, and to the Katugampola derivative \cite{Katugampola}
\begin{equation}
\mathcal{D}^\alpha (f)(x)= \lim_{\epsilon \to 0} \frac{f(x e^{\epsilon x^{-\alpha}})- f(x)}{\epsilon},
\end{equation}
that also gives $\mathcal{D}^\alpha (f)(x)= x^{1-\alpha} f'(x)$. Due to these relations, our analyses of the diffusion equation with Hausdorff derivative can be easily generalized to other kinds of local fractional derivatives.

\subsection{q-derivative}

The $q$-derivative is one of the most important deformed derivative. It is used in the Tsallis non-extensive Statistical Mechanics to stud a huge variety of complex systems. Borges proposes the $q$-derivative as follows \cite{qderivative}:
\begin{equation}
D_{(q)}f(x)\equiv \lim_{y \to x} \frac{f(x)-f(y)}{x \ominus_q y}=[1+(1-q)x]\frac{df(x)}{dx},
\end{equation}
where $\ominus$ is $x \ominus_q y \equiv \frac{x-y}{1+(1-q)y}$, $ (y \neq 1/(q-1)) $. Recently, it was show that in a first approximation, there is a relationship between the $q$-derivative and the Balankin's Hausdorff derivative with $ 1-q=\dfrac{(1-\alpha)}{l_0}$ \cite{WL}.

\section{Model description}
\label{SecM}

Let $\bar{c}=\bar{c}(x,y,z,t)$ be the average concentration of a given non-reactive contaminant in the PBL. If $\vec{u}$ is the wind velocity and $\overrightarrow{\Pi}_{c}$ is the concentration flux, then the spatial distribution of the concentration can be given by the equation \cite{Csanady, Pasquill, Seinfeld}:
\begin{equation}\label{eq 1}
 \frac{\partial \; \overline{c} }{\partial t}+\vec{u}\cdot \overrightarrow{\nabla}+\overrightarrow{\nabla}\cdot \overrightarrow{\Pi}_{c}=0.
\end{equation}
If we consider only the steady-state regime, where $\frac{\partial \; \overline{c} }{\partial t}=0$, then \eqref{eq 1} reduces to
\begin{equation}\label{eq 2}
 u\frac{{\partial}\;\overline{c}}{\partial x}+\frac{\partial}{\partial x}({\Pi}_{c,x})+\frac{\partial}{\partial y}({\Pi}_{c,y})+\frac{\partial}{\partial z}({\Pi}_{c,z})=0,
\end{equation}
where we choose a cartesian coordinate system in which the longitudinal direction $x$ coincides with the mean wind velocity, and ${\Pi}_{c,x}$, ${\Pi}_{c,y}$ and ${\Pi}_{c,z}$ are the components of the concentration flux in the directions $x$, $y$ and $z$, respectively.

In order to fix the concentration flux $\bar{c}$ from the advection-diffusion equation \eqref{eq 2}, we should state the dependence of ${\Pi}_{c,x}$, ${\Pi}_{c,y}$ and ${\Pi}_{c,z}$ with $\bar{c}$. A traditional closure for concentration flux problem \eqref{eq 2} is given by the gradient-transfer approach, which assumes that turbulence causes a net movement of material in the direction of the concentration gradient, at a rate proportional to the magnitude of the gradient \cite{Pasquill}, that is:
\begin{equation} \label{eq 2b}
\Pi_{c,y} = -K_{y} \frac{\partial \overline{c}}{\partial y}, \;\;\;\;\; \Pi_{c,z} = -K_{z} \frac{\partial \overline{c}}{\partial z},
\end{equation}
where for simplicity we neglect the longitudinal diffusion $\Pi_{c,x} =0$, as usual in the literature, and $K_{y}$ and $K_{z}$ are the eddy diffusivity.

Assuming that we use the parameterization \eqref{eq 2b} for the concentration flux equation \eqref{eq 2}, and by considering a constant value for $u$, $K_y$ and $K_z$, the solution of \eqref{eq 2} for a point source with boundary at infinity is a Gaussian distribution with a mean square displacement linear in $x$. However, we know that a diffusion process in a turbulent flow is anomalous, in the sense that the mean squared displacement is not linear \cite{Metzler2000}. A consequence from this fact is that the anomalous diffusion process obeys a non-Gaussian distribution, unlike the normal diffusion in which the distribution is Gaussian \cite{Metzler2000}. In order to deal with the anomalous diffusion introduced by turbulence in traditional atmospheric dispersion models, the velocity $u$, and the parameters $K_y$ and $K_z$, are usually considered as complex functions on $x$, $y$ and $z$. In this context, the turbulent parameterization is related to the estimate of the mean wind speed and the diffusion coefficient \cite{Moreira,Sharan}. The wind speed is obtained from the similarity theory \cite{Panofsky,Smith,Irwin} and the diffusion coefficient is obtained from the Taylor statistical diffusion theory \cite{Taylor,Batchelor}.

In the present work, we take a different approach in order to develop a model for dispersion of contaminants in the PBL. Instead of introduce complex functions for the the velocity $u$ and the parameters $K_y$ and $K_z$ to deal with anomalous diffusion, we consider an advection-diffusion equation with Hausdorff derivatives and constant $u$, $K_y$ and $K_z$. The Hausdorff derivative of order $\alpha$ was introduced in the context of a time-dependent anomalous diffusion process displaying a power-law mean square displacement with exponent $\alpha$ \cite{Hausdorff}. Furthermore, the diffusion of contaminants in the atmosphere is driven by the turbulence that exhibits a fractal structure, and the Hausdorff derivative can be easily related to the fractal dimension of the medium \cite{Balankin,Balankin2}. To introduce the fractional derivative of Hausdorff in the diffusive term we will parameterize the concentration flux as follows:
\begin{equation} \label{eq 3}
\Pi_{c,y} = -K_{y} \frac{\partial \overline{c}}{\partial y^\alpha}, \;\;\;\;\; \Pi_{c,z} = -K_{z} \frac{\partial \overline{c}}{\partial z^\alpha},
\end{equation}
where $\frac{\partial }{\partial y^\alpha}$ and $\frac{\partial }{\partial z^\alpha}$  are Hausdorff derivatives \cite{Hausdorff}, and for simplicity we neglected again the longitudinal diffusion $\Pi_{c,x}=0$, as usual in the literature. For the advective term, we replace the integer-order derivative in the variable $x$ in \eqref{eq 2} by a Hausdorff derivative in $x$. This replacement is necessary to obtain an advection-diffusion equation for an anomalous process displaying a power-law mean squared displacement \cite{Hausdorff}. Furthermore, by doing this we get the same physical dimensions on both sides of the equality \cite{Goulart2019}. In this case, the spatial distribution for the concentration $\overline{c}$ is given by the equation:
 \begin{equation}\label{eq 4}
u \frac{\partial \ \overline{c}}{\partial x^\alpha}= \frac{\partial}{\partial y} \left(K_{y}\  \frac{\partial \ \overline{c}}{\partial y^\alpha}\right)+\frac{\partial}{\partial z} \left(K_{z}\  \frac{\partial \ \overline{c}}{\partial z^\alpha}\right).
\end{equation}
Finally, in order to compare the model with experimental data found in the literature, the equation for the cross-wind integrated concentration ($\overline{c^{y}}=\overline{c^{y}}(x,z)$) is obtained by integrating the equation \eqref{eq 4} with respect to $y$ from $-\infty$ to $+\infty$:
\begin{equation}\label{eq 6a}
\frac{\partial\ \overline{c^{{y}}}}{\partial x^\alpha}= \kappa \ \frac{\partial}{\partial z} \left(  \frac{\partial \ \overline{c^{{y}}}}{\partial z^\alpha}\right),
\end{equation}
since $u$ and $K_z$ are constants, and where $\kappa=\frac{K}{u}$. Finally, by considering $\overline{c^{y}}$ differentiable, equation \eqref{eq 6a} reduces to
\begin{equation}\label{eq 6}
x^{1-\alpha}\frac{\partial\ \overline{c^{{y}}}}{\partial x}= \kappa \ \frac{\partial}{\partial z} \left(z^{1-\alpha}\frac{\partial \ \overline{c^{{y}}}}{\partial z}\right),
\end{equation}
where we use \eqref{Hdef2}.

In order to equation \eqref{eq 6} describes a possible real dispersion process in PBL, it should be imposed boundary conditions of zero flux on the ground ($z = 0 $) and top ($z = h$), and consider that the contaminant is released from an elevated point source with emission rate $Q$ at height $H_{s}$, i.e.,
\begin{equation} \label{eq 7a}
K \ \frac{\partial \overline{c^{y}} }{\partial z}=0,\;\;\;\;z=0,\;\;z=h,
\end{equation}
\begin{equation}\label{eq 8}
u \ \overline{c^{y}}(0,z)=Q\delta(z-H_{s}),\;\;x=0,
\end{equation}
where $\delta(\cdot)$ is the Dirac delta function.

The solution of the differential equation \eqref{eq 6}, subjected to the boundary conditions \eqref{eq 7a} and \eqref{eq 8}, can be analytically obtained by the separation of variables and Frobenius methods. By inserting $\overline{c^{y}}(x,z)=X(x)Z(z)$ in \eqref{eq 6} we obtain the ordinary differential equations:
\begin{equation}
\label{eq 8b}
\frac{dX}{dx}+\lambda_n^2\kappa x^{\alpha-1}X=0
\end{equation}
and
\begin{equation}
\label{eq 8c}
\frac{d^2Z}{dz^2}+(1-\alpha)z^{-1}\frac{dZ}{dz}+\lambda_n^2 z^{\alpha-1}Z=0,
\end{equation}
where $\lambda_n$ are constants to be fixed by the boundary conditions. Equation \eqref{eq 8b} can be solved by direct integration, and equation \eqref{eq 8c} can be solved with the Frobenius method by setting
\begin{equation}
\label{eq 8d}
Z(z)= \sum_{n=0}^{\infty}c_n z^{(\alpha+1)(n+r)},
\end{equation}
where $c_n$ and $r$ are constants. The solution is given by:
\begin{equation}\label{eq 9}
\overline{c^{y}}(x,z)=a_{0}+z^{\frac{\alpha}{2}}\sum_{n=1}^{\infty}a_{n}J_{-\frac{\alpha}{\alpha + 1}}\Big(\frac{2 \lambda_{n}}{\alpha + 1}z^{\frac{\alpha + 1}{2}}\Big)\text{exp} \Big(-\frac{\lambda_{n}^{2}\ \kappa \ x^{\alpha}}{\alpha }\Big),
\end{equation}
where $J_{p}(x)$ is the Bessel function of first species and order $ p $. The constants $a_{n}$ and $\lambda_{n}$ are obtained from the boundary conditions \eqref{eq 7a} and initial condition \eqref{eq 8}.

In the present work, we compare the solution \eqref{eq 9} of our model with real experimental data, and also to the Gaussian and operational Gaussian models. The Gaussian model is obtained from  equation \eqref{eq 9} taking $\alpha = 1$,
\begin{equation}\label{eq 10}
\overline{c^{y}}(x,z)=\frac{Q}{u h}\Big[1+2\sum_{n=1}^{\infty}\cos(\lambda_{n}H_{s})\cos(\lambda_{n}z)\exp (- \kappa \lambda_{n}^{2}x) \Big].
\end{equation}
A very commonly used expression in the literature for the Gaussian model is obtained from the solution of the advection-diffusion (equation \eqref{eq 9} with $\alpha=1$) in an infinite medium. For this equation to satisfy the boundary conditions given by equations \eqref{eq 7a} and \eqref{eq 8}, with $-\infty < x< \infty$ and $-\infty < z< \infty$, a 'mirror-image' source is considered. The solution for this so called operational Gaussian model (O-G model) is \cite{Csanady}
\begin{equation}\label{eq 11}
\overline{c^{y}}(x,z)=\frac{Q}{2\sqrt{\pi \kappa x }u}\Big[\exp\Big(-\frac{(z-H_{s})^{2}}{4 \kappa x}\Big)+\exp \Big(-\frac{(z+H_{s})^{2}}{4 \kappa x}\Big) \Big].
\end{equation}

In the next section we are going to compare the solution \eqref{eq 9} of our fractional model \eqref{eq 6} against experimental data and both the Gaussian \eqref{eq 10} and O-G \eqref{eq 11} solution models.

\section{Results and discussion}
\label{SecR}

In order to analyze the performance of the dispersion model with Hausdorff derivative proposed in this work, we will compare the results obtained from the \eqref{eq 9} with the classical Gaussian model \eqref{eq 10}, with the operational Gaussian model \eqref{eq 11} and with the model proposed by Goulart et al \cite{Goulart2019}, where the fractional derivative of Caputo was used in the parameterization of the concentration flux and in the advective term.  The mean wind velocity is obtained directly from the experimental data. To obtain a constant eddy diffusivity $K_z$, we follow the procedure introduced in \cite{Goulart2017,Goulart2019}, where it was considered a spatial average $K_z=\langle K\rangle$ of an eddy diffusivity that is a linear function of downwind distance expressed by  $K=\rho u x$, where $\rho$ is the turbulence parameter. The turbulence parameter $\rho$ is parameterized as the square of turbulent intensity using Taylor statistical theory of diffusion $\rho=(\frac{\sigma_{w}}{u})^{2}$ \cite{Arya}, where $\sigma_{w}$ is the standard deviation of the vertical wind speed component. The boundary layer flow can be classified as unstable (or convective), stable and neutral. In an unstable (or convective) flow the heat flux is positive relative to the surface (occurs during the day). In stable flow, the heat flux is negative relative to the surface (usually at night). A flow is neutral if the heat flow is zero or if the mechanical energy output is much higher than the thermal energy output. Regardless of this rating, as in \cite{Goulart2019} the  experimental data from the experiments of Copenhagen \cite{Gryning1984}, Prairie Grass \cite{Barad} and Hanford \cite{Nickola} were separated in two groups: one with $\frac{h}{|L|}<10$ and another with $\frac{h}{|L|}>10$, where $L$ is the Monin-Obukhov length \cite{Obukhov} and $h$ is height of PBL. The parameter $\frac{h}{|L|}$ is obtained from the energy balance equation in a turbulent flow. This parameter can be used to evaluate some characteristics of the physical structure of the turbulent flow. For $\frac{h}{|L|}<10$ we have a predominance of mechanical energy input (wind shear) in the turbulent flow. For $\frac{h}{|L|}>10$ we have a predominance of energy input by thermal convection in the turbulent flow. The analysis of these two situations aims to show that the value of $\alpha$ that best describes the concentration distribution when confronting the proposed Hausdorff derivative model with the experimental data, is related to the physical structure of the turbulent flow. Usually, the performance of dispersion models is evaluated from a well know set statistical indices described by Hanna \cite{Hanna} defined in the following way,
\begin{equation*}
\begin{split}
\text{NMSE}\; (\text{normalized mean square error})&= \frac{\overline{(c_{o}-c_{p})^{2}}}{\overline{c_{o}}\overline{c_{p}}},\\
\text{Cor}\; (\text{correlation coefficient})&= \frac{\overline{(c_{o}-\overline{c_{p}})(c_{p}-\overline{c_{p}})}}{\sigma_{o}\sigma_{p}},\\
\text{FB}\; (\text{fractional bias})&=\frac{\overline{c_{o}-\overline{c_{p}}}}{0.5(\overline{c_{o}}+\overline{c_{p}})},\\
\text{FS}\; (\text{fractional standard deviations}) &= \frac{\sigma_{o}-\sigma_{p}}{0.5(\sigma_{o}+\sigma_{p})},
\end{split}
\end{equation*}
where $c_{p}$ is the computed concentration, $c_{o}$ is the observed concentration, $\sigma_{p}$ is the computed standard deviation, $\sigma_{o}$ is the observed standard deviation, and the overbar indicates an averaged value. The statistical index FA2 represents the fraction of data for $0.5 \leq\frac{c_{p}}{c_{o}}\leq 2 $. The best results are indicated by values nearest to $0$ in NMSE, FS and FB, and nearest to $1$ in Cor and FA2.

\subsection{Results for $\frac{h}{|L|}<10$  }

To estimate the better $\alpha$ value for each experiment, we analyzed the solution of our model from $\alpha=0.50$ to $\alpha=0.99$ by steps of $0.01$. We found that, regardless of the experiment, for $\frac{h}{|L|}<10$ the Hausdorff fractional model with constant wind velocity and constant eddy diffusivity (equation \eqref{eq 9}) describes relatively well all experiments with $\alpha = 0.54$. We present separately the results of each experiment for $\frac{h}{|L|}<10$.

\subsubsection{Copenhagen experiment}

\begin{table}[!h]
\centering
\caption{\small Copenhagen Experiment for $\frac{h}{|L|}<10$ (instable)}\label{tab1}
\begin{tabular}{lccccccc}
\hline\hline
   Model                 &      Cor      &     NMSE           &                    FS             &      FB                 &    FA2        \\
\hline\hline
  Eq. \eqref{eq 9}      &      0.96     &     0.12           &                    0.10           &     -0.36               &    1.00       \\
  Goulart et al. (2019)  &      0.97     &     0.05           &                    0.08           &     -0.24               &    1.00       \\
  Gauss Model            &      0.96     &     0.17           &                    0.06           &     -0.44               &    0.75       \\
 O-G Model               &      0.97     &     0.83           &                    1.00           &     -0.77               &    0.41       \\
 Moreira (2005)          &      0.97     &     0.02           &                    0.05           &      0.01               &    1.00		\\
 Kumar (2012)            &      0.90     &     0.05           &                    0.34           &     -0.04               &    0.96       \\
  \hline\hline
\end{tabular}
\end{table}
For the Copenhagen experiment with $\frac{h}{|L|}<10$, Table \ref{tab1} shows that, when compared to the Gaussian and Operational Gaussian (O-G) models using a mean wind velocity and eddy diffusivity, our model generates good results for the concentration distribution of contaminants in the PBL generated by a turbulent flow where the source of thermal convection and mechanical input is quite relevant. It also shows a similar result to that obtained by Goulart et al \cite{Goulart2019}, especially when comparing the results for the correlation and factor of two. In the model of Goulart et al \cite{Goulart2019}, the Caputo fractional derivative was used to parameterize the concentration flux.

Our model also presents good results when compared to some Eulerian dispersion models found in the literature, which employ integer-order derivatives in the advection-diffusion equation. In the case of the Moreira model \cite{Moreira2005b}, the stationary advection-diffusion equation employs a mean wind velocity that is a function of height $z$, and an eddy diffusivity that is a function of height $z$ and horizontal distance $x$. The differential equation is solved by the GILTT method \cite{Wortmann}, extended to the case where the eddy diffusivity is a function of $z$ and $x$. In the case of the Kumar model \cite{Kumar} the GILTT method is used to solve the advection-diffusion equation, but in this case, the mean wind velocity and eddy diffusivity are functions only of the height $z$.
Table \ref{tab1} shows that the model proposed in this work employing the Hausdorff derivative, where a constant mean wind velocity and eddy diffusivity are used, has a similar performance to the model of \cite{Moreira2005b}. We also see that the fractional Hausdorff model proposed in this work performs better than Kumar model \cite{Kumar}. We can also observe that a wind speed and eddy diffusivity that more correctly describes the turbulent flow, used in the models of Moreira \cite{Moreira2005b} and Kumar \cite{Kumar}, tends to compensate the deficiency of the mathematical structure of the classical advection-diffusion equation to describe the concentration distribution \cite{Goulart2019}. In our work, a constant mean wind velocity and eddy diffusivity were used precisely to show the ability of the Hausdorff advection-diffusion equation to describes more accurately the concentration distribution of contaminants.

\begin{figure}\caption{Scatter diagram of observed and predicted concentration for the Copenhagen experiment (instable).}\label{fig1}
\includegraphics[width=0.5\textwidth]{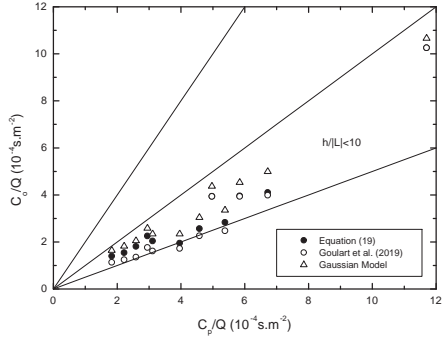}
\end{figure}

In addition to the statistical indices, Figure \ref{fig1} shows the scatter diagram of observed concentration and predicted concentration. Lines indicate a factor of two.

\subsubsection{Prairie Grass experiment}

\begin{figure}\caption{Scatter diagram of observed and predicted concentration for the Prairie Grass experiment (stable). (a) $x=200m$ \;\;  (b) $x=800m$ \;\; (c) $x\geq 200m$  \;\; (d)  $x\geq 400m$ \label{fig2}.}
\includegraphics[width=0.49\textwidth]{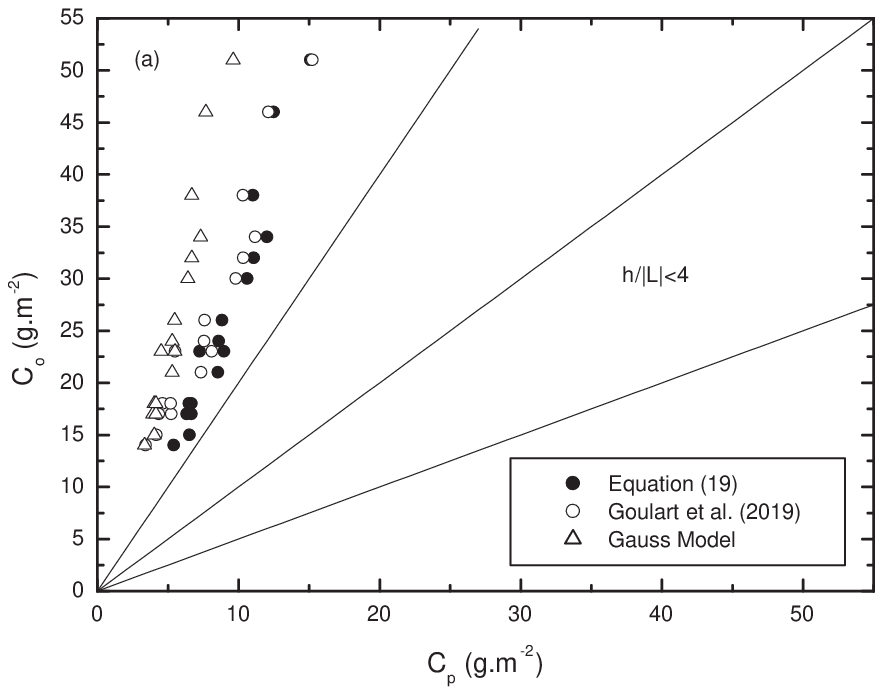}
\includegraphics[width=0.49\textwidth]{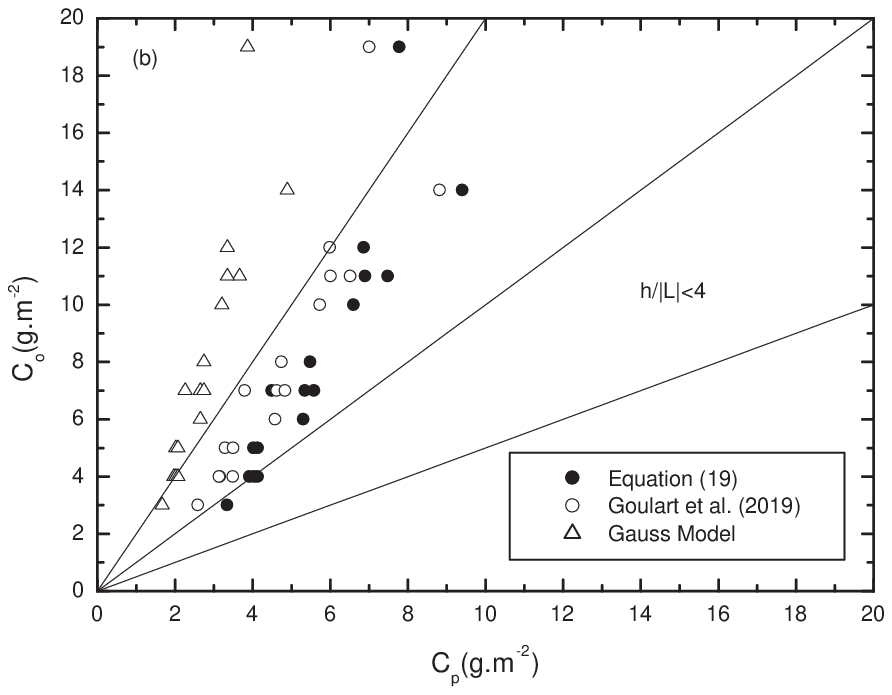}
\includegraphics[width=0.49\textwidth]{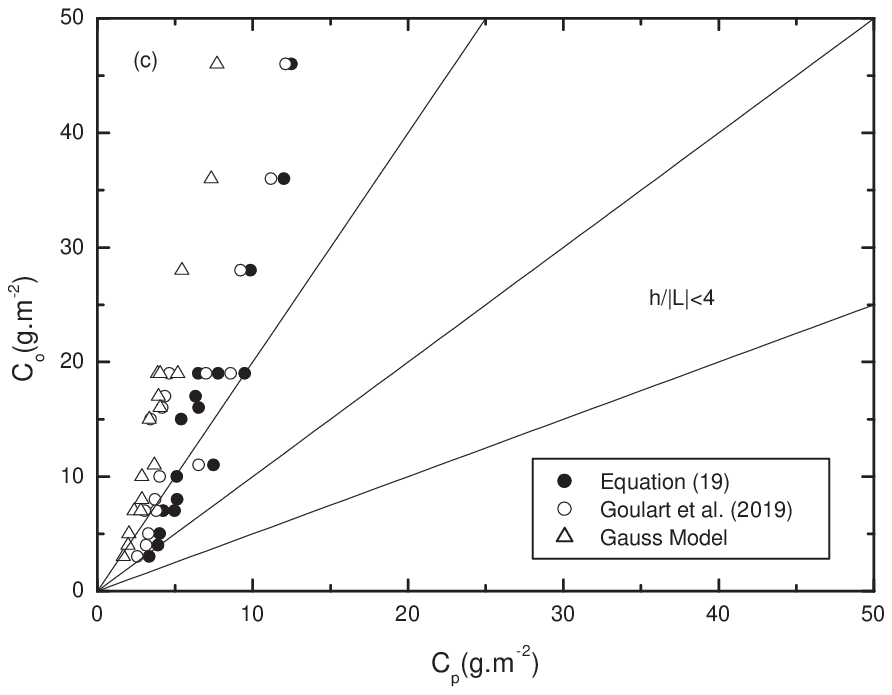}
\includegraphics[width=0.49\textwidth]{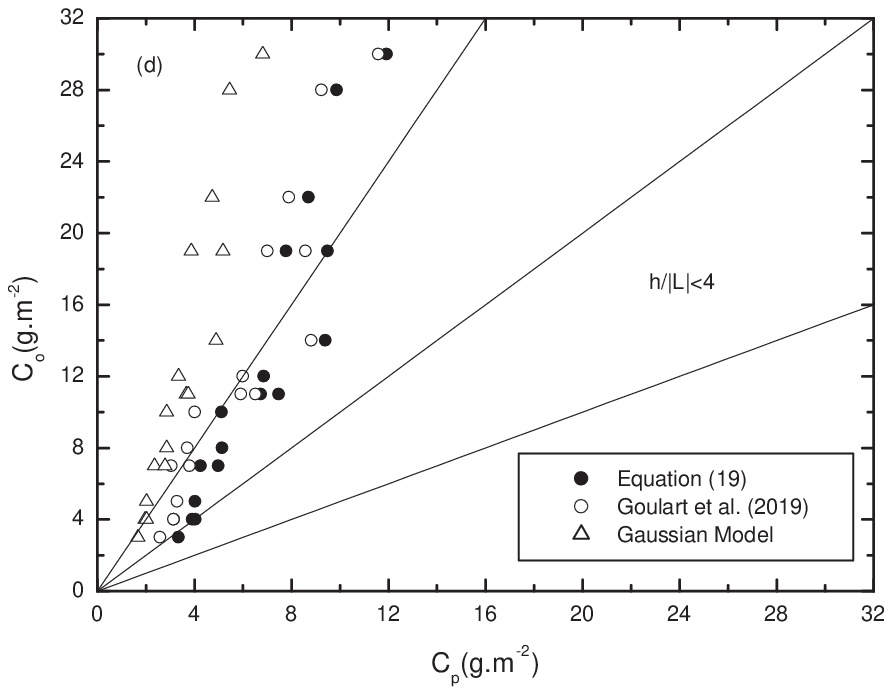}
\end{figure}

In Table \ref{tab2} we display the results obtained for the stable Prairie Grass experiment with $\frac{h}{|L|}<4$. In this case, the stability regime can be considered closer to neutral. We can observe that the model proposed in this work employing the Hausdorff derivative and the fractional model proposed by Goulart et al \cite{Goulart2019} performs better than both Gaussian and O-G models for all distances. In addition, we see that for greater distances the difference between the model proposed in this work employing the Hausdorff derivative and the fractional model proposed by Goulart et al \cite{Goulart2019} and the Gaussian model is more evident. The Figure \ref{fig2} shows the scatter diagram of observed concentration and predicted concentration for $200$m and $800$m distances.
 We attribute the better performance of our model employing the Hausdorff derivative (and also the fractional model proposed by Goulart et al \cite{Goulart2019}) in relation to the Gaussian model for greater distances to the fact that the dispersion of contaminants in a turbulent flow does not obey a Gaussian probability distribution, but rather an anomalous diffusion distribution, induced by the turbulence, with power-law mean squared displacement \cite{Metzler2000, Goulart2017, Goulart2019}. At large distances the influence of the initial conditions is small, and consequently, the distribution of probability inherent to the model is predominant in the estimation of the concentration of contaminants. We also observed that the performance of our model is slightly superior to the model proposed by Goulart et al. \cite{Goulart2019} that employs Caputo fractional derivatives.

\begin{table}[!h]
\centering
\caption{\small Prairie Grass Experiment for $\frac{h}{|L|}<4$ (stable)} \label{tab2}
\begin{tabular}{lccccccc}
\hline\hline
   Distance (m)                        &      Model                   &     Cor            &           NMSE             &      FS                &    FB        &    Fa2       \\
\hline\hline

 \multirow{3}{*}{200}                  &      Eq. \eqref{eq 9}       &     0.96           &           1.43             &     1.21               &    -0.99     &    0.00      \\
                                       &      Goulart et al. (2019)   &     0.96           &           1.68             &     1.06               &    -1.08     &    0.00      \\
                                       &      Gauss Model             &     0.96           &           3.23             &     1.47               &    -1.31     &    0.00     \\
                                       &      O-G Model               &     0.96           &           0.95             &     1.08               &    -0.83     &    0.06     \\ \hline

 \multirow{3}{*}{800}                  &      Eq. \eqref{eq 9}       &     0.90           &           0.26             &     0.87               &    -0.36     &    0.94     \\
                                       &      Goulart et al. (2019)   &     0.90           &           0.47             &     0.89               &    -0.56     &    0.88      \\
                                       &      Gauss  Model            &     0.89           &           1.56             &     1.33               &    -0.97     &    0.17     \\
                                       &      O-G  Model              &     0.84           &           1.72             &     1.35               &    -1.02     &    0.06     \\
                         \hline
  \multirow{3}{*}{$\geq 200 $}         &      Eq. \eqref{eq 9}       &     0.94           &           1.09             &     1.24               &    -0.78     &    0.53      \\
                                       &      Goulart et al. (2019)   &     0.91           &           1.35             &     1.20               &    -0.89     &    0.41      \\
                                       &      Gauss Model             &     0.97           &           2.82             &     1.46               &    -1.20     &    0.06     \\
                                       &      O-G Model               &     0.97           &           2.82             &     1.46               &    -1.20     &    0.06     \\
             \hline
  \multirow{3}{*}{$\geq 400 $}         &      Eq. \eqref{eq 9}       &     0.94           &           0.55             &     1.05               &    -0.56     &    0.79      \\
                                       &     Goulart et al. (2019)    &     0.93           &           0.73             &     1.03               &    -0.68     &    0.41      \\
                                       &      Gauss Model             &     0.94           &           2.06             &     1.40               &    -1.08     &    0.08     \\
                                       &      O-G Model               &     0.95           &           2.06             &     1.40               &    -1.08     &    0.08     \\
              \hline
  \multirow{3}{*}{All distances}       &      Eq. \eqref{eq 9}       &     0.96           &           2.91             &     1.48               &    -1.15     &    0.31      \\
                                       &      Goulart et al. (2019)   &     0.85           &           3.72             &     1.47               &    -1.27     &    0.31      \\
                                       &      Gauss Model             &     0.98           &           4.56             &     1.54               &    -1.39     &    0.02     \\
                                       &      O-G Model               &     0.98           &           4.56             &     1.54               &    -1.39     &    0.02     \\

   \hline\hline
\end{tabular}

\end{table}

\begin{table}[!h]
\centering
\caption{\small Hanford Experiment for $\frac{h}{|L|}\leq 3$ (stable)}\label{tab3}
\begin{tabular}{lccccccc}
\hline \hline
   Distance (m)                            &      Model                      &     Cor            &           NMSE             &      FS                &    FB        &    Fa2      \\  \hline \hline
  \multirow{3}{*}{100}                     &      Eq. \eqref{eq 9}          &     0.63           &           10.91            &     1.81               &    -1.73     &    0.00     \\
                                           &      Goulart et al (2019)       &     0.65           &           10.32            &     1.80               &    -1.66     &    0.00     \\
                                           &      Gauss Model                &     0.63           &           13.21            &     1.84               &    -1.71     &    0.00     \\
                                           &      O-G  Model                 &     0.62           &           15.33            &     1.88               &    -1.74     &    0.00     \\   \hline

  \multirow{3}{*}{1600}                    &      Eq. \eqref{eq 9}          &     0.98           &           0.41             &     0.68               &    -0.58     &    0.83     \\
                                           &      Goulart et al (2019)       &     0.97           &           0.36             &     0.72               &    -0.53     &    1.00     \\
                                           &      Gauss Model                &     0.97           &           0.42             &     0.69               &    -0.58     &    0.66     \\
                                           &      O-G Model                  &     0.88           &           6.21             &     1.78               &    -1.49     &    0.00     \\     \hline

  \multirow{3}{*}{3200}                     &      Eq. \eqref{eq 9}         &     0.96           &           0.11             &     0.46               &    -0.23     &    1.00     \\
                                            &      Goulart et al (2019)      &     0.95           &           0.10             &     0.50               &    -0.19     &    1.00     \\
                                            &      Gauss Model               &     0.95           &           0.12             &     0.49               &    -0.22     &    1.00     \\
                                            &      O-G Model                 &     0.84           &           6.54             &     1.81               &    -1.48     &    0.00     \\ \hline

  \multirow{3}{*}{$\leq 800 $}              &      Eq. \eqref{eq 9}         &     0.44           &           10.65            &     1.78               &    -1.59     &    0.00     \\
                                            &      Goulart et al (2019)      &     0.60           &           8.27             &     1.77               &    -1.51     &    0.05     \\
                                            &      Gauss Model               &     0.48           &           10.32            &     1.79               &    -1.58     &    0.00     \\
                                            &      O-G Model                 &     0.82           &           14.24            &     1.83               &    -1.71     &    0.00     \\   \hline

  \multirow{3}{*}{$\geq 1600 $}             &      Eq. \eqref{eq 9}         &     0.92           &           0.28             &     0.63               &    -0.42     &    0.92     \\
                                            &      Goulart et al (2019)      &     0.92           &           0.26             &     0.67               &    -0.38     &    1.00 \\
                                            &      Gauss Model               &     0.90           &           0.29             &     0.65               &    -0.42     &    0.83     \\
                                            &      O-G Model                 &     0.83           &           6.31             &     1.75               &    -1.49     &    0.00     \\    \hline

  \multirow{3}{*}{All distances}            &      Eq. \eqref{eq 9}         &     0.35           &           7.58             &     1.75               &    -1.42     &    0.37     \\
                                            &      Goulart et al (2019)      &     0.55           &           7.57             &     1.74               &    -1.35     &    0.43     \\
                                            &      Gauss Model               &     0.39           &           9.31             &     1.76               &    -1.42     &    0.33     \\
                                            &      O-G Model                 &     0.87           &           14.01            &     1.80               &    -1.68     &    0.00     \\   \hline \hline
\end{tabular}
\end{table}

\begin{figure}\caption{Scatter diagram of observed and predicted concentration for the Hanford experiment. (a) $x=1600m$ \;\; (b) $x=3200m$  \;\; (c)  $ x\geq 1600m $ \label{fig3}}
\includegraphics[width=0.49\textwidth]{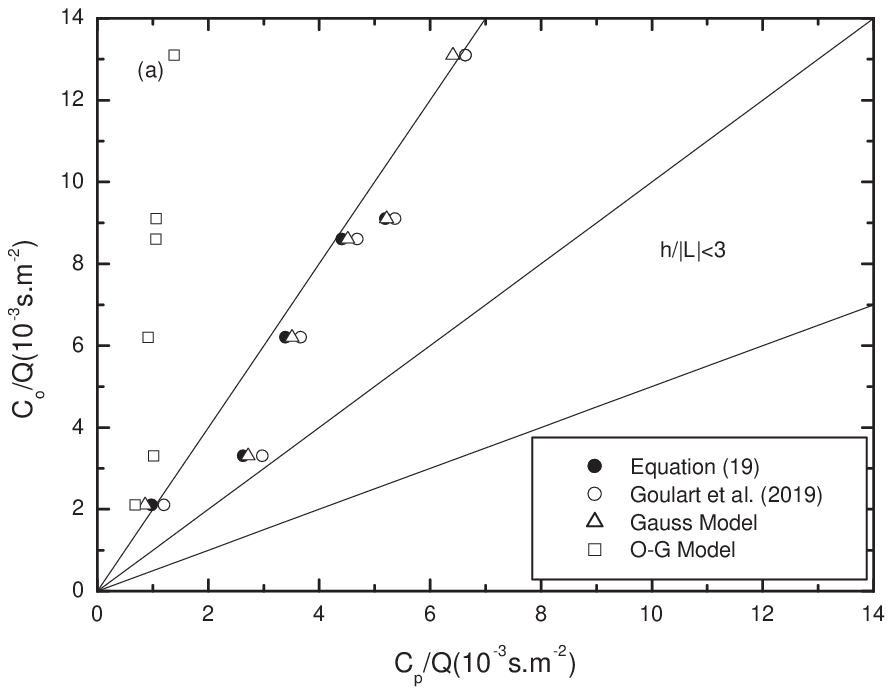}
\includegraphics[width=0.49\textwidth]{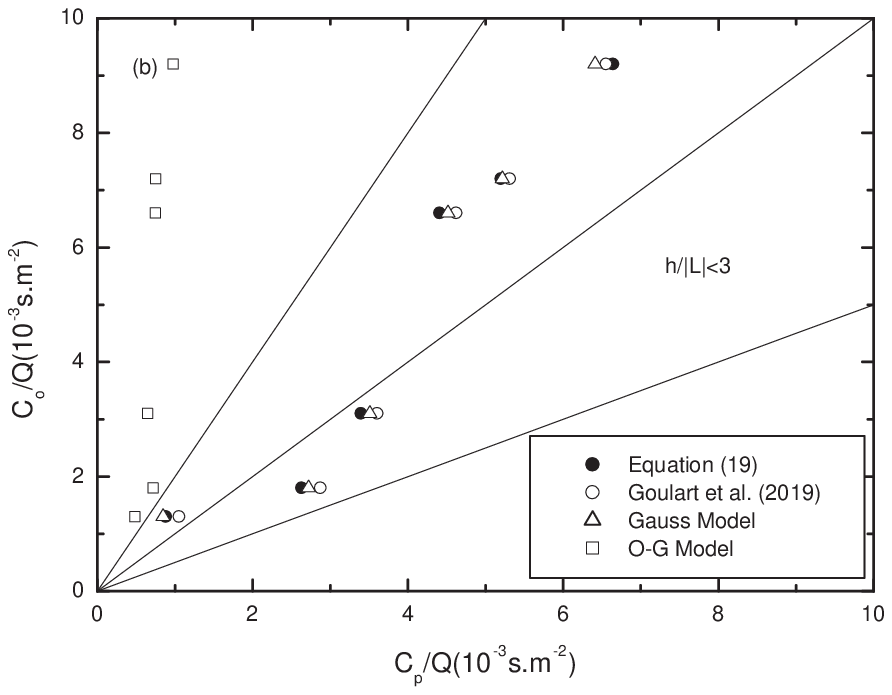}
\includegraphics[width=0.49\textwidth]{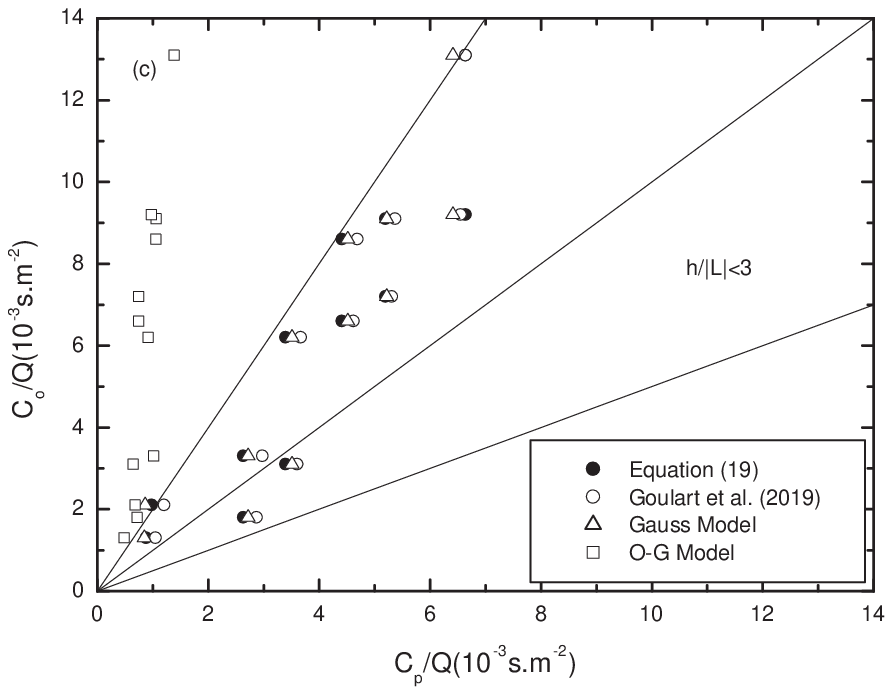}
\end{figure}

\subsubsection{Hanford experiment}

Table \ref{tab3} and Figure \ref{fig3} shows the results of our model, and also the fractional \cite{Goulart2019}, the Gaussian and the O-G models, for the Hanford experiment with $\frac{h}{|L|}<3$. We can verify that, for small distances, the results obtained with all these models are very bad and, for large distances, the Gaussian model, the fractional model \cite{Goulart2019} and the model proposed in this work employing the Hausdorff derivative, have similar performances. Especially for a distance of $3200$m from the source. The O-G model fails completely at all distances. In the Hanford experiment, we have a stable regime, with very low wind speed ($\approx 1.5 m s^{-1}$). Therefore, the flow is slightly turbulent and the characteristics of anomalous diffusion, present in a strong turbulent flow, are not very evident. In this case, it is expected that the Gaussian probability distribution can be employed to model the problem. We attribute the low performance of the O-G model to the approximation made so that the solution satisfies the boundary conditions \eqref{eq 8} and \eqref{eq 9}. The values of the distances in the $z$ and $x$ directions involved in the experiments of Copenhagen and Prairie Grass are similar in almost all experiments. In this case, the approximation of the boundary at infinity gives a good result. In particular, for the stable Prairie Grass where the $z$ and $x$ lengths are $\approx 400$m and $800$m, respectively, the difference between Gaussian and O-G is negligible. In the Hanford experiment, the values of the distances in the $z$ and $x$ directions are very different ($ \approx 200$m and $3200$m, respectively), making the boundary at infinity approximation totally inadequate.

\subsection{Results for $\frac{h}{|L|}>10$ }

As in the previous case, we analyzed the solutions of our model from $\alpha=0.60$ to $\alpha=0.99$, by steps of $0.01$, and we found that our model (with Hausdorff derivatives and constant mean wind velocity and constant eddy diffusivity) describes relatively well all experiments with $\alpha = 0.80$ when $\frac{h}{|L|} > 10$.

\begin{table}[!h]
\centering
\caption{\small Copenhagen Experiment for $\frac{h}{|L|}>10$ (instable)}\label{tab4}
\begin{tabular}{lccccccc}
\hline \hline
   Model                &      Cor      &     NMSE           &                    FS             &      FB                 &    FA2        \\
\hline \hline
  Eq. \eqref{eq 9}     &      0.62     &     0.30           &                    1.08           &     -0.29               &    0.91       \\
Goulart et al. (2019)   &      0.65     &     0.20           &                    0.97           &     -0.14               &    0.90       \\
  Gauss Model           &      0.62     &     0.34           &                    1.02           &     -0.33               &    0.81       \\
  O-G Model             &      0.97     &     0.83           &                    1.00           &     -0.77               &    0.41       \\
 \hline \hline
\end{tabular}
\end{table}

\begin{table}[!h]
\centering
\caption{\small Prairie Grass Experiment for $\frac{h}{|L|}>10$ (instable)}\label{tab5}
\begin{tabular}{lccccccc}
\hline \hline
  Distance (m)         &      Model                &     Cor            &           NMSE             &      FS                &    FB       &    Fa2      \\
\hline \hline
  \multirow{3}{*}{50}  &      Eq. \eqref{eq 9}     &     0.87           &           0.45             &     -0.76              &    0.60     &    0.75     \\
                       &      Goulart et al. (2019)&     0.80           &           0.37             &     -1.04              &    0.07     &    0.75     \\
                       &      Gauss                &     0.89           &           0.40             &     -0.76              &    0.58     &    0.75     \\
                       &     O-G Model             &     0.89           &           0.40             &     -0.76              &    0.58     &    0.75     \\ \hline
  \multirow{3}{*}{100} &      Eq. \eqref{eq 9}     &     0.47           &           1.40             &     -1.40              &    0.96     &    0.15     \\
                       &      Goulart et al. (2019)&     0.48           &           1.26             &     -1.59              &    0.66     &    0.65     \\
                     &      Gauss                  &     0.41           &           1.24             &     -1.38              &    0.91     &    0.20     \\
                     &      O-G Model              &     0.41           &           1.24             &     -1.38              &    0.91     &    0.20     \\
  \hline
\end{tabular}
\end{table}

The Tables \ref{tab4} and \ref{tab5} shows the results for the experiments of Copenhagen and Prairie Grass $\frac{h}{|L|} > 10$. We also observed that the performance of all models for experiments with $\frac{h}{|L|} > 10$ is worst than the performance when $\frac{h}{|L|} < 10$ (Tables \ref{tab1}, \ref{tab2} and \ref{tab3}). This may be related to the difference in the structure of the turbulent flow in the PBL in both cases. When $\frac{h}{|L|} > 10$ a free-convection-like state emerges \cite{Wyngaard}.  In this case, we have strong turbulence generated by thermal convection and a large variation in the structure of the flow in a vertical direction. This implies a large variation with the height of the intensity of the vertical eddy diffusivity. As in the experiments of Prairie Grass (unstable), and cases 1, 3, 7 and 8 of the Copenhagen experiment. On the other hand, when $\frac{h}{|L|}$ is very small ($\frac{h}{|L|} < 10$) there is turbulence where the mechanical source (wind shear) is relevant. In this case, we have a greater spatial homogenization in the flow. Consequently, we have a low variation with the height of the intensity of the vertical eddy diffusivity \cite{Plein}, as in the Prairie Grass (stable), and the Hanford experiments, and also in cases 2, 4, 5, 6 and 9 of the Copenhagen experiment.
Models that employ a constant wind speed and eddy diffusivity will have greater difficulty in describing correctly the concentration distribution when there is a large spatial asymmetry, as in the case of the PBL flow if $\frac{h}{|L|} > 10$. ThisIn this present work deficiency will be naturally compensated when non-constant wind speed and eddy diffusivity are used to more accurately describe the physical structure of the flow. However, in the present work we consider only constant wind speed and eddy diffusivity since, at this moment, we are interested only in demonstrating that a differential equation with Hausdorff derivatives is more appropriate to calculate the distribution of contaminants in turbulent flow than a differential equation with classical derivatives. We are confronting a model with deformed derivatives against the Gaussian model with classical integer-order derivatives. Of course, both non-classical or classical models which consider a non-constant wind speed and eddy diffusivity, that more adequately describes the structure of the turbulent flow, will generate more realistic concentration distribution values. However, in general, it is not simple to obtain an analytic solution for both equations to easily compare its performances.


\section{Conclusions}
\label{SecC}

Our main purpose in the present work was to investigate the potential of a diffusion equation with local non-classical derivatives to model real data for the steady-state spatial concentration of contaminants in the atmosphere. The modeling of the dispersion of pollutants in the atmosphere is a challenging problem due to the turbulence of the airflow. The multi-scaling behavior of a turbulent medium, related to its fractal structure \cite{Mandelbrot}, has as a consequence the emergence of the anomalous diffusion phenomenon that plays a central role in the dispersion of contaminants in the PBL. A diffusion process is considered anomalous if it displays a non-linear time-dependent mean square displacement, instead of a linear one that characterizes a common diffusive process \cite{Metzler2000}. Moreover, from this fractal behavior, it is expected that the classical advection-diffusion equation does not describe adequately the dispersion of pollutants in the atmosphere since the parameters of the system usually grow faster than the solution obtained from the classical advection-diffusion equation \cite{West,West2}.

In this scenario, the use of fractional derivatives to model anomalous diffusion emerged as a valuable mathematical tool \cite{Metzler2000}. Moreover, the promising results found in our previous works \cite{Goulart2017,Goulart2019}, by using Caputo fractional derivatives to model the steady-state spatial concentration of contaminants in the PBL, motivate a deeper analysis of the relationship between turbulent flow and anomalous diffusion expressed mathematically by fractional diffusion equations. In this context, in the present work, we propose an advection-diffusion equation with Hausdorff derivatives. The use of Hausdorff derivatives has two advantages against Caputo in the analyses of a turbulent anomalous diffusion process. The first is that they are local operators, resulting in a diffusion equation easier to solve. The second advantage is that they can be easily related to the fractal dimension of the medium \cite{Balankin,Balankin2}. In order to investigate the potential of a diffusion equation with Hausdorff derivative, we compare the solution of our model with real data for the steady-state spatial concentration of contaminants in the atmosphere. Furthermore, we also compare the present model with the solutions from the Caputo fractional derivative model \cite{Goulart2019}, and also to integer-order models.

The results obtained show that an advection-diffusion equation with Hausdorff derivatives gives better results than classical integer-order derivatives models when fitting real data. Actually, the models with Hausdorff and Caputo fractional derivatives have similar performance when compared to experiments. However, the Hausdorff has the practical advantage of being a local operator instead of the non-local Caputo operator. Consequently, a diffusion equation with Hausdorff derivatives is easier to be generalized and solved than a Caputo equation when the wind speed and eddy diffusivity are complex functions describing more adequately the real system. Furthermore, this result indicates that regardless of the kind of non-classical derivative we use, an advection-diffusion equation with non-classical derivative displaying power-law mean square displacement is more adequate to describe the diffusion of contaminants in the atmosphere than a model with classical derivative. Moreover, since Hausdorff derivatives can be related to several deformed derivatives \cite{WL,W}, our result highlights the potential of deformed derivatives models to describe the diffusion of contaminants in the atmosphere. Finally, a very important result we found it is that there should be a relation between the order $\alpha$ of the Hausdorff fractional derivative with the physical structure of the turbulent flow since, regardless the experiment, when we have a predominance of mechanical energy input in the turbulent flow all experiments are better described with $\alpha = 0.54$, and when we have a predominance of energy input by thermal convection the experimental data is better described by $\alpha = 0.80$. For future works, it should be important to obtain a direct relationship between the fractal structure of the turbulent flow in real experiments with the order $\alpha$ of the fractional Hausdorff derivative in the model.


\section*{Acknowledgments}

This work was supported in part by CNPq and CAPES, Brazilian funding agencies.

\appendix


\end{document}